\documentclass{PoS}
\usepackage{bm}
\usepackage{amsmath}
\usepackage{makerobust}
\MakeRobustCommand\eqref

\title{Nuclear Parity Violation from Lattice QCD}

\ShortTitle{Nuclear Parity Violation}

\author{\speaker{Thorsten Kurth}\thanks{LLNL-PROC-678488}\\
        Nuclear Science Division, Lawrence Berkeley National Laboratory\\
        Department of Physics, University of California, Berkeley\\
        E-mail: \email{tkurth@lbl.gov}}
\author{Evan Berkowitz, Enrico Rinaldi, Pavlos Vranas\\
	Nuclear and Chemical Sciences Division, Lawrence Livermore National Laboratory\\
	E-mail: \email{berkowitz2@llnl.gov}, \email{rinaldi2@llnl.gov}, \email{vranas2@llnl.gov}}
\author{Amy Nicholson, Mark Strother\\
	Department of Physics, University of California, Berkeley\\
	E-mail: \email{anicholson@berkeley.edu}, \email{mcstro@berkeley.edu}}
\author{Andre Walker-Loud\\
	Nuclear Science Division, Lawrence Berkeley National Laboratory\\
	Theory Center, Thomas Jefferson National Accelerator Facility\\
	Department of Physics, The College of William \& Mary\\
	E-mail: \email{awalker-loud@lbl.gov}}

\abstract{The electroweak interaction at the level of quarks and gluons are well understood from precision measurements in high energy collider experiments.
Relating these fundamental parameters to Hadronic Parity Violation in nuclei however remains an outstanding theoretical challenge. One of the most interesting observables in this respect is the parity violating hadronic neutral current: it is hard to measure in collider experiments and is thus the least constrained observable of the Standard Model. 
Precision measurements of parity violating transitions in nuclei can help to improve these constraints.
In these systems however, the weak interaction is masked by effects of the seven orders of magnitude stronger non-perturbative strong interaction.
Therefore, in order to relate experimental measurements of the parity violating pion-nucleon couplings to the fundamental Lagrangian of the SM, these non-perturbative effects have to be well understood. In this paper, we are going to present a Lattice QCD approach for computing the $\Delta I{=}2$ parity violating matrix element in proton proton scattering. 
This process does not involve disconnected diagrams in the isospin symmetric limit and is thus a perfect testbed for studying the feasibility of the more involved calculation of the parity violating pion-nucleon coupling.}

\FullConference{The 33rd International Symposium on Lattice Field Theory\\
		14 -18 July 2015\\
		Kobe International Conference Center, Kobe, Japan*}

\begin{document}

\section{Introduction}
It is well known that parity is conserved by QCD and QED but violated by the weak interaction. After its discovery in beta decays of $^{60}\mathrm{Co}$ by Wu et. al., subsequent measurements at collider experiments have determined the charged-current interaction to very good precision. With new data from the LHC and the discovery of the Higgs boson~\cite{Aad:2012tfa-Chatrchyan:2012ufa}, almost every corner of the Standard Model has been tested to high precision through combined experimental and theoretical efforts. The least constrained observable of the Standard Model today is the neutral hadronic current
\begin{equation}
J_\mu^0\equiv \bar{u}\gamma_\mu(1-\gamma_5)u-\bar{d}\gamma_\mu(1-\gamma_5)d-4\sin^2\theta_WJ_\mu^{\mathrm{em}},
\end{equation}
where $\theta_W$ is the Weinberg angle and $J_\mu^\mathrm{em}$ the electromagnetic current. Flavor conservation renders it difficult to study the neutral hadronic current in collider experiments. Nuclear systems on the other hand are perfect testbeds for performing measurements of hadronic parity violation (PV). This is due to the fact that the charged current suppresses nuclear isospin transitions of $\Delta I{=}1$ with respect to $\Delta I{=}0,2$ processes by a factor of $\sin^2\theta_C{\approx 0.04}$, where $\theta_C$ is the Cabibbo angle. In neutral current decays however, this suppression is absent and the $\Delta I{=}1$ channel therefore provides a unique opportunity to study the weak neutral current. The quantitative study of this channel would allow us to determine if the parity violating force is long range, since single-pion exchange is solely responsible for $\Delta I{=}1$ transitions.

A series of experiments have been performed aiming at measuring hadronic PV in nuclear systems. At these energy scales, the non-perturbative nature of QCD masks these operators due to their weak intrinsic scale of $G_FF_\pi^2{\sim} 10^{-7}$, where, $F_\pi{\sim}92.4\,\mathrm{MeV}$ is the pion decay constant and $G_F$ the Fermi constant. 
The daunting task of measuring asymmetries to one part in $10^7-10^8$ led experimenters to first search for PV effects in nuclei with enhanced sensitivity due to nearly degenerate opposite parity states, resulting in findings of asymmetries of a few percent~\cite{Krane:1971zz-Krane:1971zza-Alfimenkov:1983-Masuda:1988mb-Yuan:1991zz}. While encouraging, the many body nuclear effects prohibit a direct connection with the underlying theory. A number of experiments utilizing longitudinally polarized protons on unpolarized proton targets $\vec{p}p$ have been performed finding non-zero results~\cite{Nagle:1978vn-Balzer:1980dn-Balzer:1985au-Kistryn:1987tq-Eversheim:1991tg}. These experiments are more ideal for connecting the results to the fundamental theory but they require precision greater than $10^{-7}$, a challenging task.

The most recent attempt is made by the NPDGamma collaboration at the Spallation Neutron Source at Oak Ridge National Laboratory. This experiment measures asymmetries in the final state  photon distribution in the capture of cold polarized neutrons on parahydrogen ($np\rightarrow d\gamma$) and aims at a sensitivity of $10^{-8}$. It has the potential to improve the results of previous attempts~\cite{Cavaignac:1977uk-Gericke:2011zz} which failed to find non-zero results. Successful measurements of this matrix element could help to reduce systematic uncertainties for experimental measurements of PV in larger nuclei as well. Nevertheless, a good understanding of the strong dynamics is of great importance and ab initio numerical methods such as Lattice QCD can help to significantly improve theoretical constraints on the hadronic neutral current.\\

\section{Details of the Calculation}
At the fundamental level, the weak interaction is mediated by the charged and neutral gauge bosons $W^\pm$ and $Z$ respectively.  At low energies however, these interactions can be treated as effective local four-fermion operators with a $V{-}A$-type coupling.
Although the most interesting processes occur in the isovector channel, we restrict ourselves to the isotensor channel, i.e. $\Delta I{=}2$. The corresponding effective operator is given by
\begin{equation}\label{eq:deltaI2op}
\mathcal{O}^{\Delta I{=}2}(t)=-\sum\limits_\mathbf{x}\left[(\bar{q}\gamma_\mu\gamma_5\tau^3q)(\bar{q}\gamma_\mu\tau^3q)-\frac{1}{3}(\bar{q}\gamma_\mu\gamma_5\vec{\tau}q)(\bar{q}\gamma_\mu\vec{\tau}q)\right](\mathbf{x},t).
\end{equation}
In contrast to the operators mediating $\Delta I{=}0,1$ transitions, it can easily be checked that this operator does not contain flavor diagonal parts and thus does not involve quark-line disconnected diagrams. Furthermore, it can be shown that under the absence of QED, the operator (\ref{eq:deltaI2op}) does not mix with its $\Delta I{=}0,1$ counterparts under renormalization~\cite{Tiburzi:2012hx}.

The process we are going to consider is PV in proton-proton scattering, i.e. the following matrix element:
\begin{equation}
\left\langle pp(\,^1\!S_0)\middle| \mathcal{O}^{\Delta I{=}2} \middle| pp(\,^3\!P_0) \right\rangle.
\end{equation}
In our lattice calculation, we employ techniques derived in~\cite{Basak:2005ir,Dudek:2010wm,Luu:2011ep} for creating sources and sink operators which have overlap with states of desired spin and parity. 
We will describe the source construction in more detail in the following. For our single nucleon operators $N$ we use the local operator labelled $G_{1g}^{1}$ in~\cite{Basak:2005ir} which was shown to have a good overlap with the single nucleon ground state in~\cite{Basak:2007kj}. We additionally apply gaussian smearing to increase the overlap with the single nucleon ground state. Our two-proton operators $pp$ can be generally written as
\begin{align}
&pp^{Jm_J}_{Im_I;S\ell}(|\bm{\Delta x}|) =
\sum\limits_{\substack{m_S,m_{\ell} \\ m_{s_1},m_{s_2} \\ m_{I_1},m_{I_2}}}
\mathrm{CG}^{Jm_J}_{\ell m_{\ell},Sm_S}
\mathrm{CG}^{Sm_S}_{s_2m_{s_2},s_1m_{s_1}}
\mathrm{CG}^{1,1}_{\frac{1}{2}m_{I_1},\frac{1}{2}m_{I_2}}
\sum\limits_{\substack{R \in \mathcal{O}_h\\\mathbf{x}}} Y_{\ell m_{\ell}}(\widehat{R \bm{\Delta x}})~ 
N_{m_{s_1}}^{m_{I_1}}(\mathbf{x}) N_{m_{s_2}}^{m_{I_2}}(\mathbf{x}+R \bm{\Delta x}),
\end{align}
where the factors $\mathrm{CG}$ denote Clebsch-Gordan coefficients which project our operator to the correct spin, isospin ($I{=}m_I{=}1$ in our case) and angular momentum quantum numbers. For projecting our operator to the desired orbital angular momentum, we employ spherical harmonics $Y_{\ell m_\ell}$, restricted to the lattice sites at a given $R\bm{\Delta x}$.

On the lattice, rotational invariance is broken and therefore angular momentum $J$ is not a good quantum number. Instead, we 
project our operators to good irreducible representations of the cubic group. Starting from operators with the continuum angular momentum labels, this can be achieved by applying the so-called subduction matrices derived in~\cite{Dudek:2010wm}.
Using those, we project our initial and final states to the cubic irreps $A_1^+$ and $A_1^-$, which have good overlap with the desired states $^1\!S_0$ and $^3\!P_0$ respectively. These and other operators were already used in~\cite{Berkowitz:2015eaa} for successfully computing two-nucleon scattering phase shifts in higher partial waves.

\begin{figure}[ht]
\centering
\includegraphics[scale=0.4]{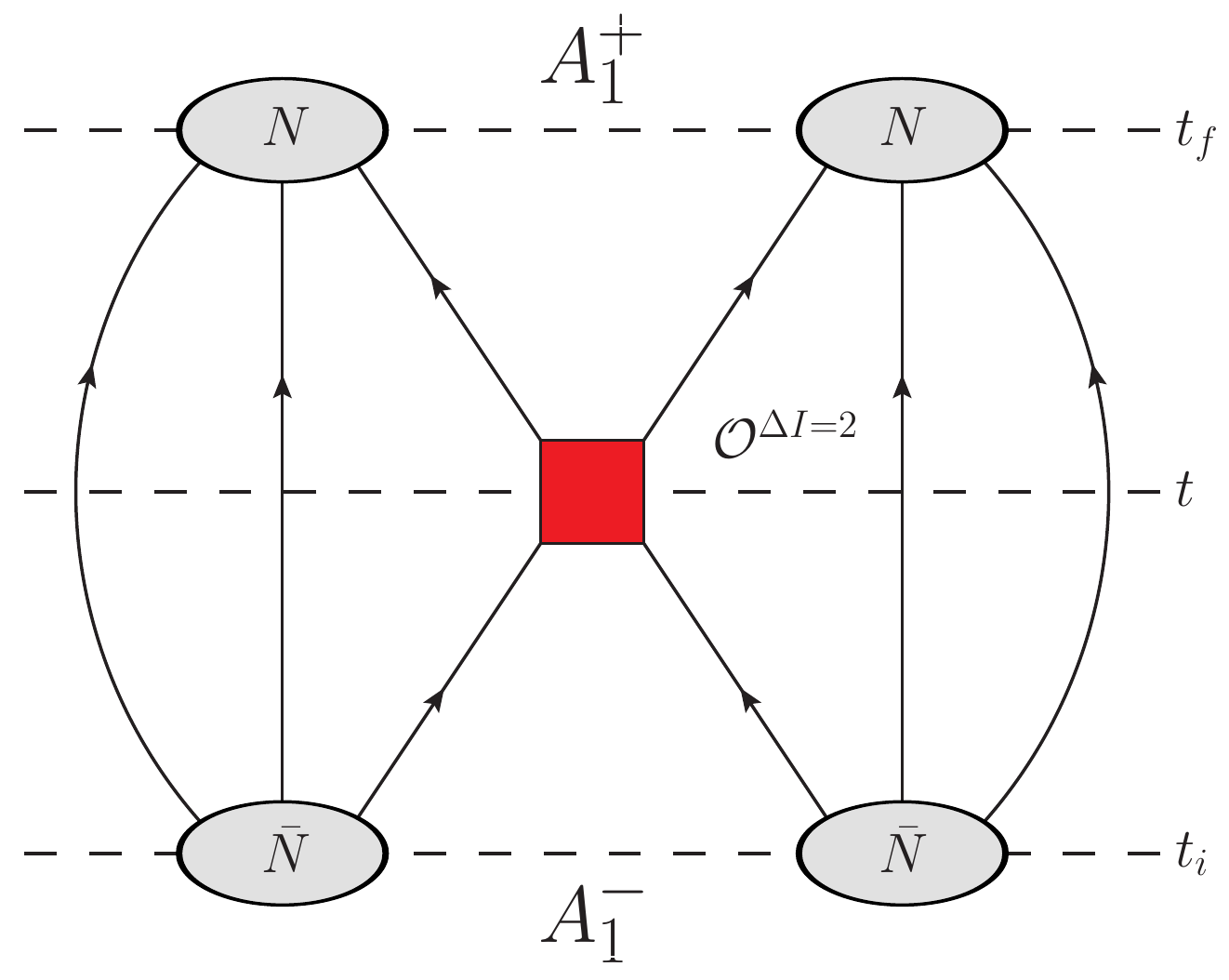}
\caption{\label{fig:contractions} Graphical description of~\eqref{eq:fivepointcorr}: the blobs denote single nucleon interpolating operators, relatively displaced by $\bm{\Delta x}$. The red square is the PV effective operator~\eqref{eq:deltaI2op} and the lines correspond to quark propagators. The diagram shown is one out of 2208 contributing Wick contractions.}
\end{figure}

We leave source and sink time slices $t_i$ and $t_f$ fixed and vary the time slice $t$ at which we insert $\mathcal{O}^{\Delta I{=}2}$. Thus, the quantity we actually compute is\footnote{We also computed the $nn\rightarrow pp$, $\Delta I{=}1$ parity violating amplitude which is related to (\ref{eq:fivepointcorr}) by isospin rotation. We found good agreement between our two results.}
\begin{equation}\label{eq:fivepointcorr}
C_{+-}(t_f,t,t_i)=\left\langle pp(A_1^+)(t_f)\middle| \mathcal{O}^{\Delta I{=}2}(t) \middle| pp(A_1^-)(t_i) \right\rangle.
\end{equation}
Figure~\ref{fig:contractions} depicts this correlation function along with a typical Wick contraction.

\begin{figure}[ht]
\centering
\includegraphics[width=0.48\textwidth]{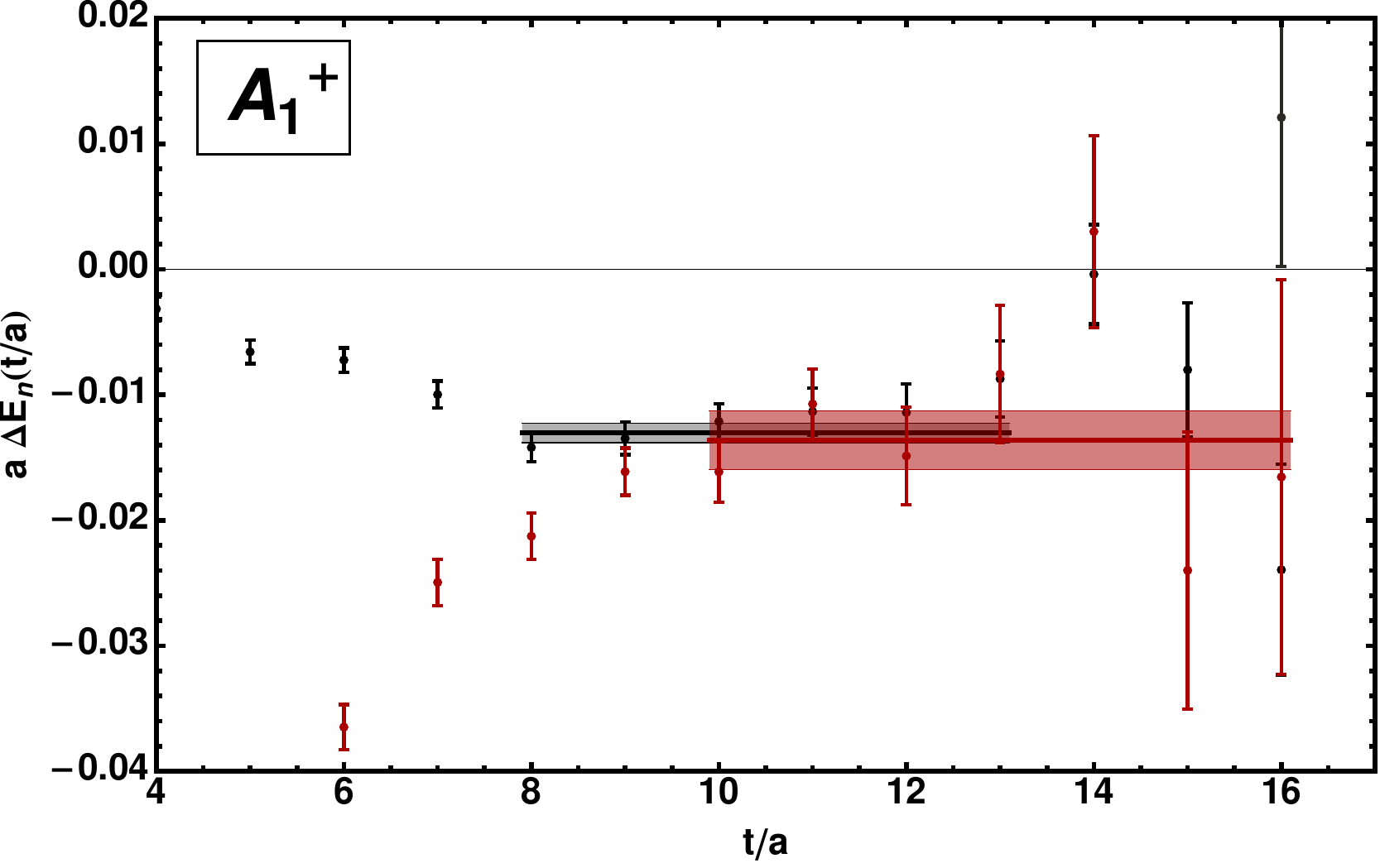}\hfill
\includegraphics[width=0.48\textwidth]{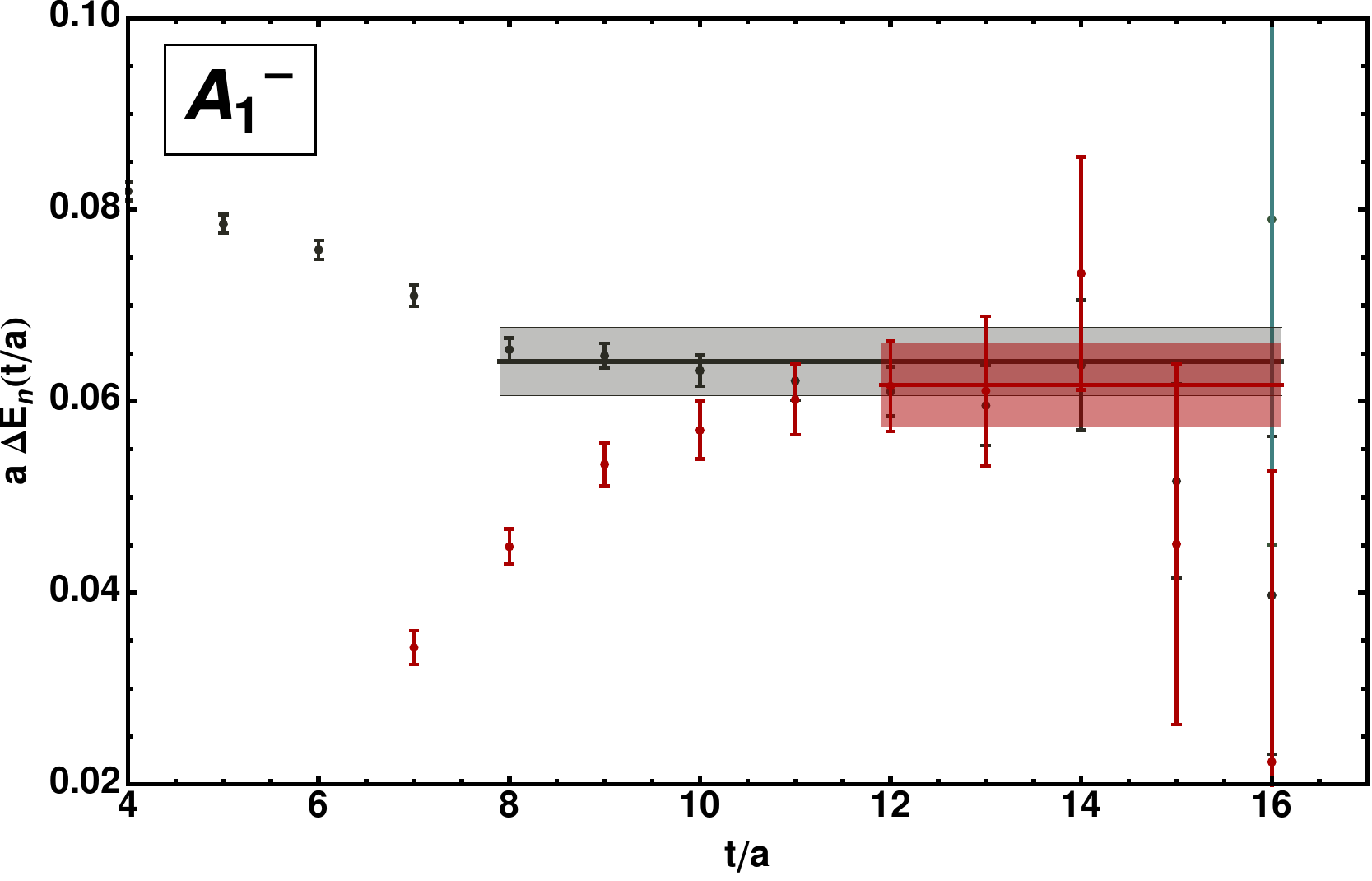}
\caption{\label{fig:irrepoverlaps} Effective masses for the ground state in $A_1^+$ (left) and $A_1^-$ channel (right). The black points and bands correspond to coordinate-to-momentum space correlation functions, whereas the red points and bands correspond to coordinate-to-coordinate space correlation functions.}
\end{figure}

When source and sink separations are sufficiently large, we expect~(\ref{eq:fivepointcorr}) to approximately plateau for values of $t$ which satisfy $t_i\ll t \ll t_f$, up to exponential corrections in the energy difference between in- and out states. This dependence as well as the dependence on the interpolating operators used can be almost completely eliminated by further considering the following amplitudes
\begin{align}\label{eq:pc_scattering}
C_{\pm}(t)&=\left\langle pp(A_1^\pm)(t) \middle| pp(A_1^\pm)(0) \right\rangle
\end{align}
and forming the ratio (cf. e.g.~\cite{Wasem:2011tp})
\begin{equation}\label{eq:rratio}
R_{+-}(t_f,t,t_i)\equiv \frac{C_{+-}(t_f,t,t_i)}{\sqrt{C_{++}(t_f-t_i)\,C_{--}(t_f-t_i)}}\sqrt{\frac{C_{++}(t_f-t)\,C_{--}(t-t_i)}{C_{--}(t_f-t)\,C_{++}(t-t_i)}}.
\end{equation}
For finite lattice spacing $a$, operator~(\ref{eq:deltaI2op}) injects energy into the system. Correlation function~(\ref{eq:rratio}) receives, at leading order, corrections proportional to $E_{A_{1}^-}-E_{A_1^+}$~(cf. e.g. \cite{Beane:2002ca}), which is approximately the energy difference between the $P$- and $S$-wave states. In order to subtract this contribution to leading order, we also compute $C_{-+}(t_f,t,t_i)$, i.e. the process with incoming $A_1^+$ and outgoing $A_1^-$ states, and compute the difference
\begin{equation}\label{eq:rratiosub}
R(t_f,t,t_i)\equiv\frac{1}{2}\big(R_{-+}(t_f,t,t_i)-R_{+-}(t_f,t,t_i)\big).
\end{equation}
We have to discuss the following important subtlety associated with our choice of operators and the setup of the lattice calculation: optimal two-nucleon operators which have a good overlap with lattice irreps $A_1^+$ and $A_1^-$ have sparse support only in momentum space. This would require us to perform an exact momentum projection at the source and sink. On the other hand, the operator has to be projected to zero momentum. If we would like to achieve both, we would have to compute all-to-all propagators. In order to avoid that costly computation, we employ coordinate space sources and sinks and perform an exact momentum projection at the operator insertion. Therefore, we have to show that our coordinate space operators have good overlap with the irreducible representations in question. This is demonstrated in Figure~\ref{fig:irrepoverlaps}. Black dots and bands denote data and fits obtained from the coordinate space to momentum space operators used in~\cite{Berkowitz:2015eaa}. The red points and bands on the other hand denote coordinate space to coordinate space operators used in this study. We observe a good agreement in the energy levels and are thus confident that our operators have good overlap with the desired states. For the operators in this paper, we use spatial displacements $\bm{\Delta x}=\Delta\cdot(1 1 1)$ and all rotations allowed by $O_h$. We found that $\Delta{=}6$ gave the best results, i.e. they maximized the overlap with the corresponding $A_1^\pm$ ground states.

\section{Results}
\begin{figure}[ht]
\centering
\includegraphics[width=0.55\textwidth]{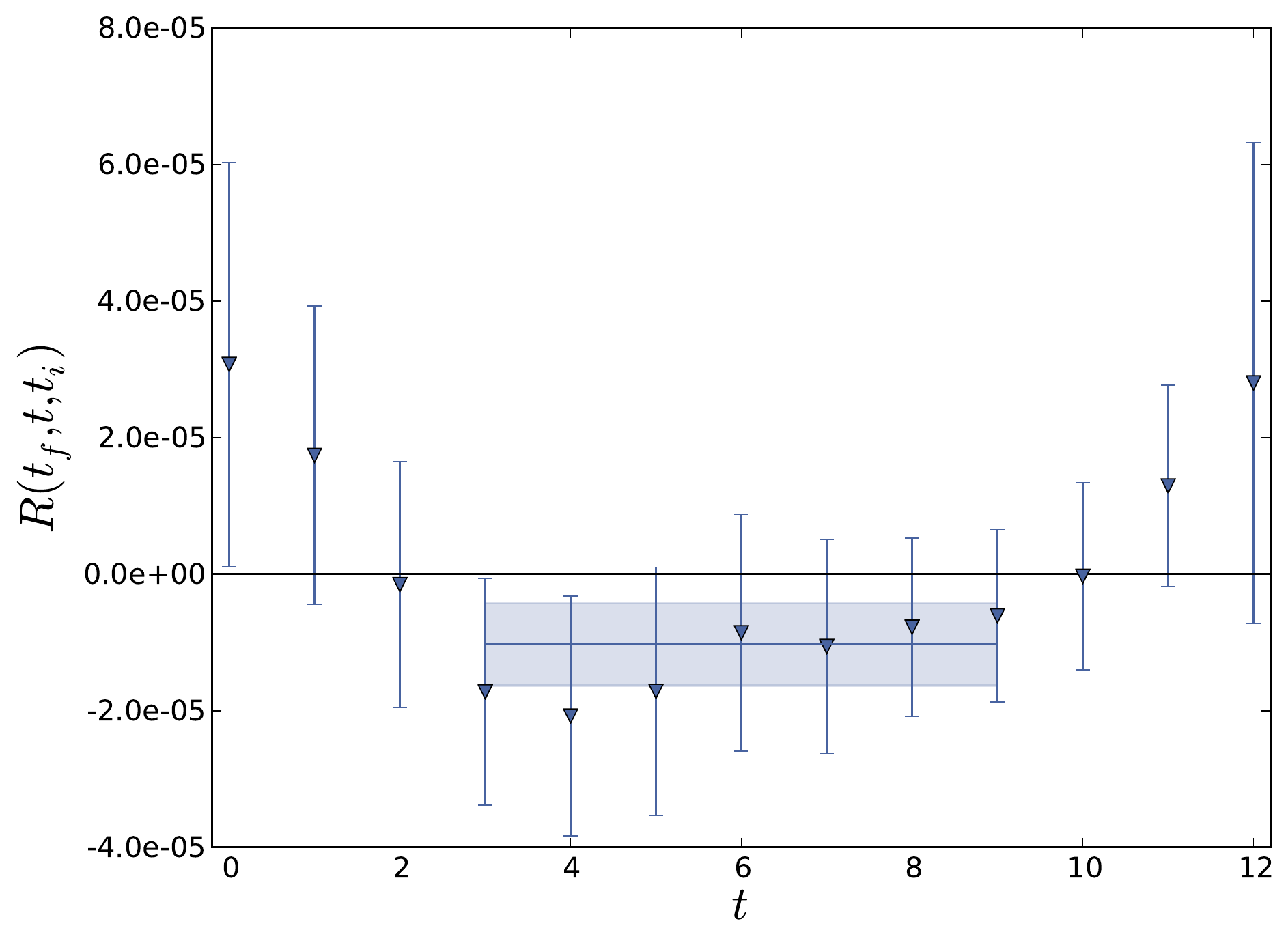}
\caption{\label{fig:ratioplot} Bare finite volume matrix element $\langle pp(A_1^-)| \mathcal{O}^{\Delta I{=}2}| pp(A_1^+)\rangle_\mathrm{FV}$. The horizontal line below zero corresponds to a constant fit and the error band was obtained by computing the standard deviation over results for the same fit obtained from 2000 bootstrap samples.}
\end{figure}

For our lattice calculation we employed isotropic Wilson clover lattices with $a{\sim}0.145\,\mathrm{fm}$, $m_\pi{\sim}800\,\mathrm{MeV}$ and $V{=}24^3{\times}48$~(c.f. \cite{Beane:2012vq}). We performed in total ${\sim}6400$ measurements on ${\sim}320$ configurations.
The parity conserving scattering amplitudes in (\ref{eq:pc_scattering}) were computed at the same time in order to preserve correlations between the numerator and denominator in (\ref{eq:rratio}). The subtracted ratio (\ref{eq:rratiosub}) is shown in Figure~\ref{fig:ratioplot}. The result for the fit is
\begin{equation}\label{eq:matelement}
M_\mathrm{FV}^{\Delta I{=}2}=-1.0(7)\cdot 10^{-5}.
\end{equation}
The error bar is purely statistical and was determined by computing the variance of the results to a constant fit on 2000 bootstrap samples. Note that we have not determined the overall normalization as well as applied the renormalization factor computed in~\cite{Tiburzi:2012xx} yet.
In this calculation, we kept the distance between source time-slice $t_i$ and sink time-slice $t_f$ fixed to $12$. We also started exploratory calculations for $t_f-t_i{=}25$ and obtained compatible results but a larger error bar due to increasing noise.

\begin{figure}[ht]
\centering
\includegraphics[width=0.48\textwidth]{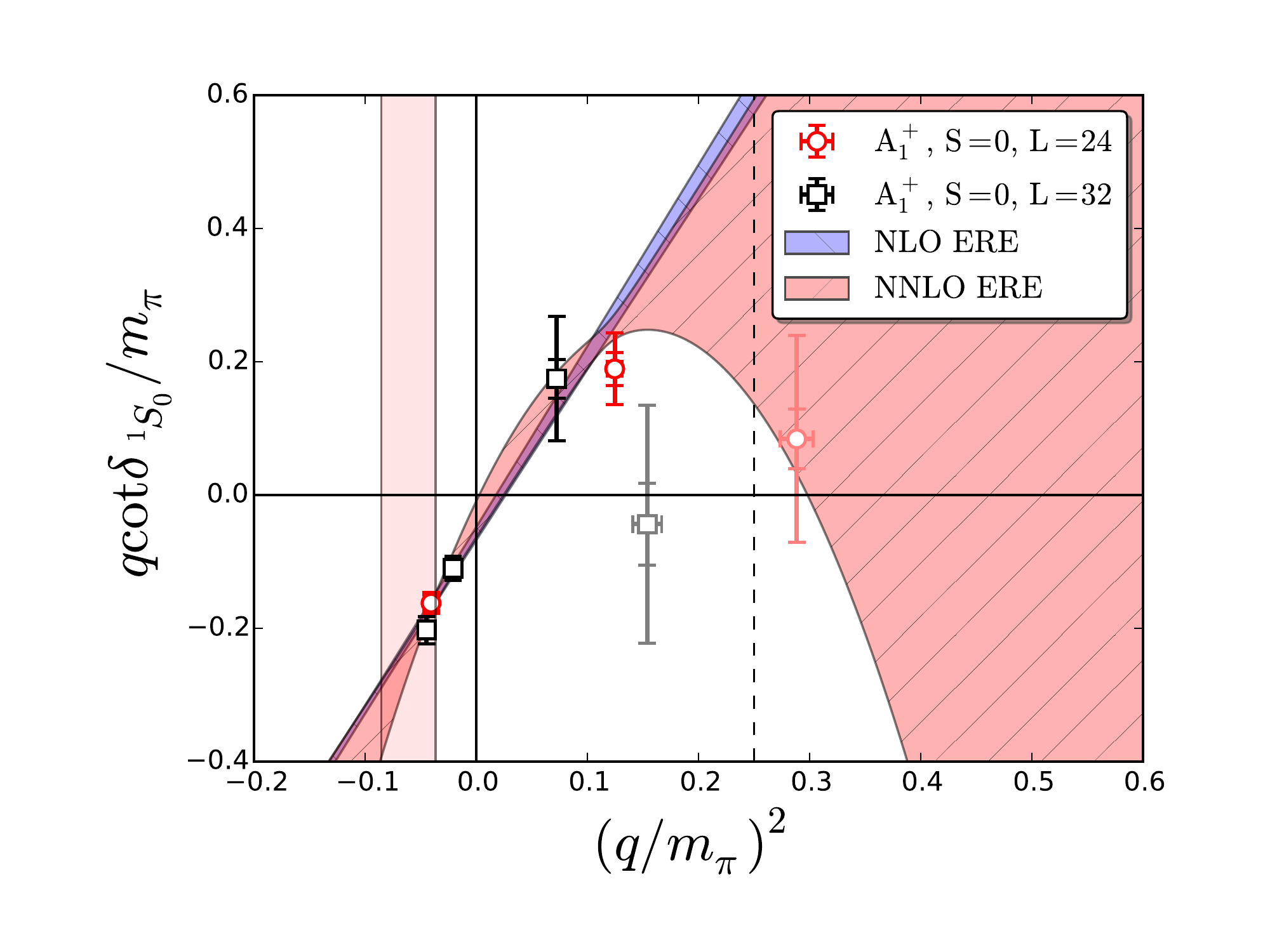}\hfill
\includegraphics[width=0.48\textwidth]{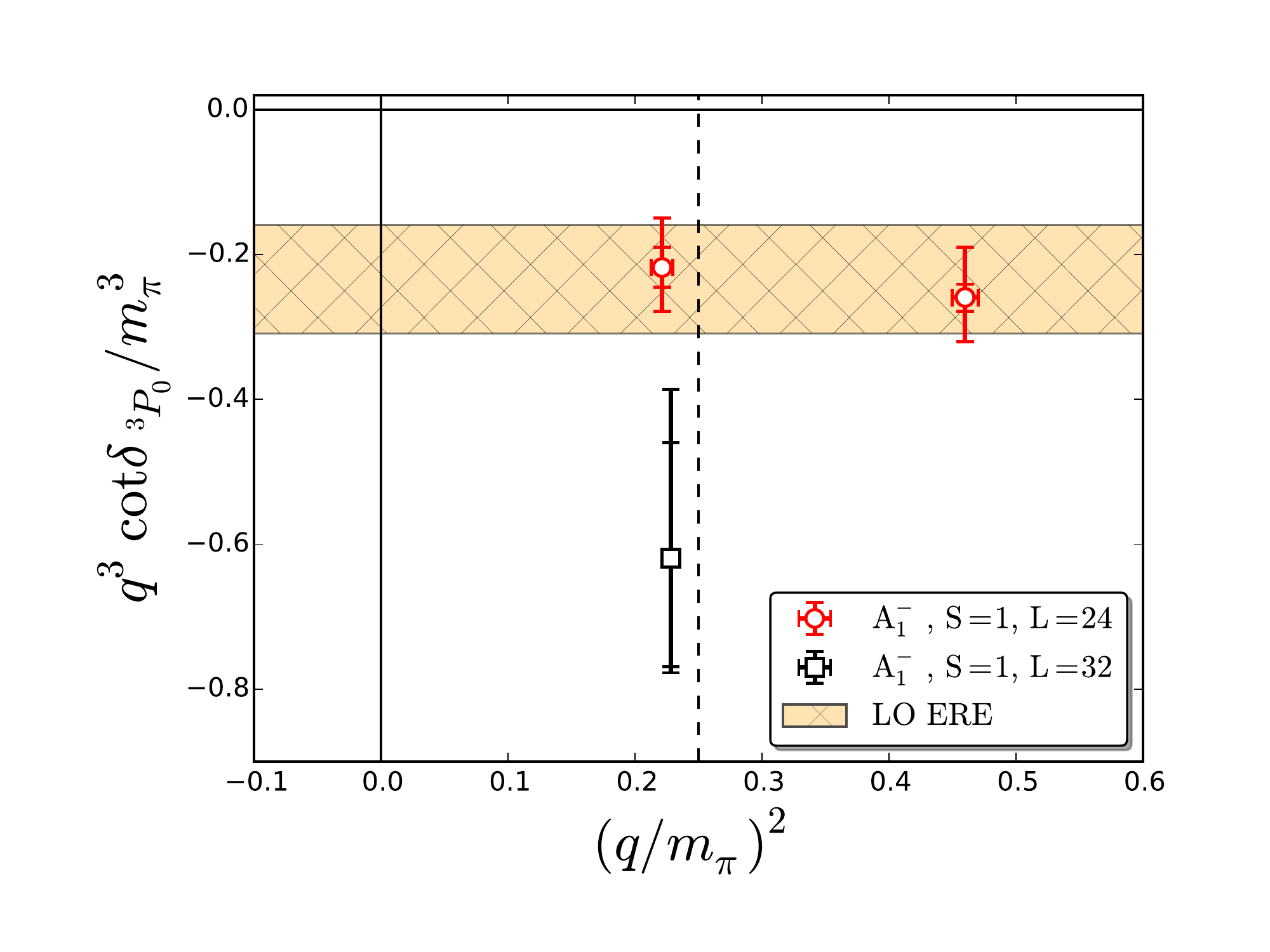}
\caption{\label{fig:phaseshifts} Energy dependence of phase shifts in $A_1^+$ (left) and $A_1^-$ channels (right). The open circles and squares correspond to phase shifts obtained by L\"uscher's finite volume method on lattices with $L{=}24$ and $L{=}32$ respectively. The shaded bands correspond to error bands obtained by performing a fit to the effective range expansion in NLO and NNLO (left) and LO (right). The dashed vertical line represents the t-channel cut beyond which the convergence of the effective range expansion is not guaranteed. Note however, that L\"uscher's finite volume method is valid even beyond that cut and our fit to the effective range expansion shows reasonable behavior even beyond the t-channel cut.}
\end{figure}

In the future, we want to relate the infinite volume matrix element (\ref{eq:matelement}) to its infinite volume counterpart. For that purpose, we need to compute the Lellouch-L\"uscher matching factor~\cite{Lellouch:2000pv} which generally depends on the phase shifts of all involved partial waves as well as their derivatives with respect to energy. Thus, we need to compute the phase shifts $\delta_{^1\!S_0}$ and $\delta_{^3\!P_0}$ and their energy dependence to a high precision.
Preliminary results for phase shifts in the $A_1^+$ and $A_1^-$ channels, which predominantly couple to the physical $^1\!S_0$ and $^3\!P_0$ channels respectively, are shown in Figure~\ref{fig:phaseshifts}.\footnote{The technology we employed to compute these phase shifts has been discussed in~\cite{Berkowitz:2015eaa}.} The open circles and squares were obtained by L\"uscher's finite volume method~\cite{Luscher:1986pf-Luscher:1990ux} on lattices with $L{=}24$ and $L{=}32$ respectively. The shaded bands correspond to fits to the effective range expansion to various orders and the dashed vertical line denotes the t-channel cut. We observe that the energy dependence of $\delta_{A_1^+}$ is mapped out to a sufficient precision. In case of $\delta_{A_1^-}$ we only obtained three data points with relatively large error bars. Thus, depending on the actual form of the yet unknown matching factor, we might need to increase our statistics or even add more points by performing the calculation on larger volumes.

\section{Summary}
We presented the preliminary results for the first lattice calculation of the parity violating $\Delta I{=}2$ nuclear matrix element. We have obtained a result for the unnormalized, bare matrix element which is significantly different from zero. We are confident that our setup is well suited for computing this quantity to a precision of a few percent. Future efforts will focus on increasing statistics, perform the renormalization as well as the infinite volume matching. For that purpose, we need to calculate the Lellouch-L\"uscher matching factor and probably increase statistics for the $A_1^-$ phase shift measurement.


\end{document}